\documentclass[]{aa}
\usepackage{graphicx}
\usepackage{epsfig}
\usepackage{txfonts}
\usepackage{natbib}
\bibpunct{(}{)}{;}{a}{}{,}

\def\Msol{\thinspace\hbox{$\hbox{M}_{\odot}$}}

\def\ie{{\it i.e.} }                    
\def\a4{\hsize 17.0cm \vsize 25.cm}

\newcommand{\der}[2]  { \frac{{\rm d}#1}{{\rm d}#2} }

\begin{document}
\title{Super stellar clusters with a bimodal hydrodynamic solution: an 
Approximate Analytic Approach}

\titlerunning{Super stellar clusters with a bimodal hydrodynamic solution}

\author{R. W\"unsch \inst{1} \and S. Silich \inst{2} \and J. Palou\v{s}
\inst{1} \and G. Tenorio-Tagle \inst{2}}

\offprints{R. W\"unsch}

\institute{Astronomical Institute, Academy of Sciences of the Czech
Republic, v.v.i., Bo\v{c}n\'\i\ II 1401, 141 31 Prague, Czech Republic
\and
Instituto Nacional de Astrof\'isica Optica y Electr\'onica, AP 51, 
72000 Puebla, Mexico}

\date{Received 12 February 2007 / Accepted 22 May 2007}

\abstract
{}
{We look for  a simple analytic model  to distinguish between stellar clusters
undergoing a bimodal hydrodynamic solution from those able to  drive only a
stationary wind. Clusters in the bimodal regime undergo strong radiative cooling
within their  densest inner regions, which results in the accumulation of the
matter injected by supernovae and stellar winds and eventually in the formation
of further stellar generations, while their outer regions sustain a stationary
wind.}
{The analytic formulae are derived from the basic hydrodynamic equations. Our
main assumption, that the density at the star cluster surface scales almost
linearly with that at the stagnation radius, is based on results from
semi-analytic and full numerical calculations.}
{The analytic formulation allows for the  determination of the threshold
mechanical luminosity that separates clusters evolving in either of the two
solutions. It is possible to fix the stagnation radius by simple analytic
expressions and thus to determine the fractions of the deposited matter that
clusters evolving in the bimodal regime blow out as a wind or recycle into
further stellar generations.}
{}


\maketitle


\section{Introduction}

The discovery of young massive stellar clusters or super stellar clusters
(hereafter SSCs) with typical masses between several times $10^4$\Msol \, to 
several times $10^6$\Msol \, and radii in the range $1 - 10$~pc has unveiled  
the dominant mode of star formation in starburst and interacting galaxies 
(see, for example, \citealt{2003ApJ...596..240M}; 
\citealt{2006astro.ph.12695W} and
references therein). SSCs dominate the structure of the ISM in their host
galaxies via their large UV photon output and their high velocity outflows or
star cluster winds. Furthermore, the interaction of the winds from nearby SSC
may lead to the inner structure of the large-scale galactic outflows,
or super galactic winds, which connect starburst events with the intergalactic
medium as in M82 \citep{2003ApJ...597..279T}.

Here, we center our attention on the hydrodynamics of the matter
reinserted within the SSC volume, particularly in the case of massive and
compact SSCs. The latter have been shown to be exposed to the strong
radiative cooling (see \citealt{2004ApJ...610..226S}) that leads to a bimodal
hydrodynamic solution \citep{2006astro.ph.12184T}. This phenomenon occurs
because the cluster mechanical luminosity and the injected gas density
are nearly linear functions of the cluster mass, while the cooling rate is
proportional to the square of the density, and therefore if one considers very
massive and compact clusters, radiative cooling would always dominate (see
\citealt{2003ApJ...590..791S} for details). Radiative cooling depletes the
energy in the densest inner regions enclosed within  the so-called {\em
stagnation radius}. The gas velocity at the stagnation radius is
zero km s$^{-1}$. Thus the matter reinserted through stellar winds and
supernovae within the central volume is accumulated to  become thermally
and/or gravitationally unstable leading to further episodes of star
formation \citep{2005ApJ...628L..13T}. 
On the other hand, the matter injected in the outer layers, between the
stagnation radius and the cluster surface, composes a stationary wind. In the
adiabatic solution and for low mass and extended clusters, the stagnation 
radius
is always found at the star cluster center. However, if radiative cooling is
taken into consideration, it progressively moves towards the star cluster
surface as one considers  more massive and compact clusters. 

In this work, as in \citet{1985Natur.317...44C}, the density, 
temperature and velocity are functions of the radial coordinate only. 
Such an approach does not take into account the effects of the turbulence 
generated by individual stellar winds and supernovae within the cluster
volume. By changing the density, velocity and temperature distributions,
it is thought to provide an additional transport of energy towards the 
smallest condensations. However it cannot prevent catastrophic cooling  
within the densest parts of the flow because it results from the net energy 
balance inside the cluster.

The properties of such clusters are described here by an approximate analytical
formulation, in agreement with our former semi-analytic and numerical
calculations (see sections 2 and 3). We display our results in a series of
universal diagrams valid for all massive and compact SSCs. Section 4 deals with
the impact of different heating efficiencies. 

\section{Adiabatic vs radiative solutions}

The original framework to study the fate of the matter deposited by supernovae
and strong stellar winds within a star cluster volume was developed by
\citet{1985Natur.317...44C}. In their adiabatic model the kinetic energy
released by individual sources is completely thermalized in situ  via direct
collisions of the high velocity gaseous streams from neighboring  sources. This
results in a high central temperature ($\sim 10^7$ - $10^8$K) and
pressure that exceeds the pressure of the surrounding
interstellar medium. The velocity of the injected matter then acquires a
particular distribution, increasing  from zero km s$^{-1}$ at the star cluster
center to the speed of sound ($c_{SC}$) at the star cluster surface, approaching
rapidly a supersonic value ($V_{A\infty}$) as it moves outside the cluster volume.
$c_{SC}$ is defined as: $c_{SC}= [(\gamma -1)/(\gamma + 1)]^{1/2} V_{A\infty}$;
where $V_{A\infty} = [2 L_{SC} / {\dot M}_{SC}]^{1/2}$ is the adiabatic wind
terminal velocity, $L_{SC}$ is the star cluster mechanical luminosity, and
${\dot M}_{SC}$ is the rate of mass injection provided by supernovae and stellar
winds. The condition for a stationary flow is then:

\begin{equation}
      \label{statcond}
{\dot M}_{SC} = 4 \pi R^2_{SC} \rho_{SC} c_{SC} , 
\end{equation}
where $R_{SC}$ is the radius of the cluster and $\rho_{SC}$ is the density 
of the outflow at the star cluster surface. The relation holds whenever 
$\rho_{SC}$ reaches the proper value, which is in direct proportion to the 
density at the stagnation radius (\ie in the center), $\rho_{st}$: 
$\rho_{SC} =  [(\gamma+1)/(6\gamma+2)]^{(3\gamma+1)/(5\gamma+1)} \rho_{st}$,
where $\rho_{st}$ is \citep{2000ApJ...536..896C}
\begin{equation}
      \label{eq.1}
\rho_{st} = \left(\frac{\gamma+1}{\gamma-1}\right)^{1/2}
\left(\frac{6\gamma+2}{\gamma+1}\right)^{\frac{3\gamma+1}{5\gamma+1}}
\frac{q_m R_{SC}}{3 V_{A\infty}} .
\end{equation}
Here $q_m = (3 {\dot M}_{SC}) / (4 \pi R^3_{SC})$ is the mass deposition rate
per unit volume which for a given $R_{SC}$ scales linearly with the star
cluster mass, $M_{SC}$, and $\gamma$ is the ratio of specific heats. Note that in the  adiabatic solution,
if one fixes the radius of the cluster, $R_{SC}$, and the adiabatic terminal
velocity, $V_{A\infty}$, the central density and the density at the star
cluster surface grow linearly with the mass of the considered cluster, since the
stagnation radius is located at the star cluster center and thus the density
profiles remain self-similar.

It has recently been shown, however, that the adiabatic model is
inadequate for very massive and compact star clusters, as in these cases 
radiative cooling becomes a dominant factor \citep{2004ApJ...610..226S}. 
The authors demonstrated that in the radiative case the wind density 
at the stagnation radius changes to
\begin{equation}
      \label{rhoc}
\rho_{st} = \mu_{i} q^{1/2}_m \left[\frac{\eta V^2_{A\infty}/2 - 
           c^2_{st} / (\gamma - 1)}{\Lambda(T_{st},Z)}\right]^{1/2} , 
\end{equation}
where $\mu_{i} = 14/11 m_H$ is the mean mass per ion ($m_H$ is the proton
mass), $\Lambda(T_{st},Z)$ is the cooling function (the cooling rate is $Q= n_e
n_i \Lambda(T_{st},Z)$), $T_{st}$ and $c_{st}$ are the temperature and speed
of sound at the stagnation radius, and $Z$ is the metallicity of
the plasma. The parameter $\eta$ describes how much of the deposited
energy is transformed into energy of the outflow (see \S4 for a detailed
discussion). We assume $\eta = 1$ everywhere in \S2 and \S3.

For clusters of a given size, the density at the stagnation radius grows in all
cases with the mass of the cluster. In the case of low mass clusters it results
from a combination of two effects: i) the mass deposition rate per unit volume,
$q_{m}$, is a linear function of $\dot{M}_{SC} (\sim M_{SC})$. ii) the
temperature at the stagnation radius, $T_{st}$, decreases with increasing
cluster mass $M_{SC}$ making $c_{st}$ smaller and thus the term in brackets in
the rhs of Eq.~(\ref{rhoc}) larger. The combined action of these two factors
then causes that the radiative solution proposed by \citet{2004ApJ...610..226S}
is almost identical to the adiabatic solution of \citet{1985Natur.317...44C}.

However, if one considers more massive clusters (or clusters with a higher
mechanical luminosity, $L_{SC}$) one would soon reach a critical value for which
strong radiative cooling would set in at the star cluster center, drastically
changing the inner wind structure (see \citealt{2006astro.ph.12184T}). For
clusters above this limit, the temperature $T_{st}$ at the stagnation radius
$R_{st}$ does not change any more and remains equal to the value that
corresponds to the highest possible pressure there (see section \ref{pmaxsec}
below). Thus the second term in Eq.~(\ref{rhoc}) remains fixed and the density
at the stagnation radius, $\rho_{st}$, grows as the square root of the star
cluster mass, whereas the mass deposition rate continues to grow linearly with
the mass of the considered cluster. This implies that above the threshold line,
the stationary condition Eq.~(\ref{statcond}) cannot be fulfilled unless the
stagnation radius moves towards the star cluster surface.
\citet{2006astro.ph.12184T} showed how to extend the semi-analytic approach
proposed in \citet{2004ApJ...610..226S} for this case. The authors also found
the semi-analytic solution for the outflow beyond the stagnation radius, and
compared it with full numerical calculations performed with the finite
difference hydrodynamic code ZEUS for which the cooling routine was modified to
make it suitable for the modeling of extremely fast cooling regions. In
particular, they include the cooling rate in the computation of the time-step
assuming that the amount of energy that can be radiated away from a given cell
during one time-step must be smaller than 10\% of its internal energy. The
time-step is decreased to meet this cooling rate condition. However, since this
could lead to extremely small time-steps, which would substantially degrade the
overall code performance, they do not allow the {\em global} time-step to
decrease below 0.1 times the "hydrodynamic" time-step determined by the
Courant-Friedrich-Levi criterion. If a certain cell requires an even smaller
time-step due to the cooling rate condition, the time-step was subdivided even
further and then was used to numerically integrate the energy equation only in
the affected cell(s). This "time refinement" was applied only {\em locally}, so
the CPU time is not wasted in cells where the high time resolution is not
required (see \citealt{2006astro.ph.12184T} for more details).

\section{The catastrophic cooling regime}

\subsection{The temperature at the stagnation radius}
\label{pmaxsec}

Eq.~(\ref{rhoc}) allows one to determine the pressure at the stagnation
radius $P_{st} = \rho_{st}k T_{st}/\mu$ where $k$ is the Boltzmann
constant, and $\mu = 14/23 m_{H}$ is the mean mass per particle.
\citet{2006astro.ph.12184T} showed that in the bimodal regime the pressure at
the stagnation radius always attains the  maximum value allowed by the shape
of the cooling function, and thus one can find the corresponding temperature
numerically if $V_{A\infty}$ and the shape of the cooling function,
$\Lambda(T,Z)$, are known. 
\begin{figure}
\centering
\resizebox{\columnwidth}{!}{\includegraphics{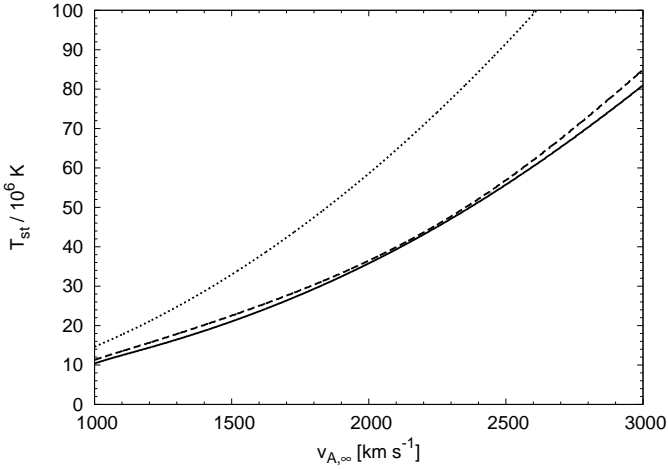}}
\caption{The temperature at the stagnation radius as a function of the
adiabatic wind terminal speed, $V_{A\infty} = 2 L_{SC} / {\dot M}_{SC}$.
Solid and dashed lines represent the temperature at the stagnation radius
for solar and 10 times solar metallicities, respectively. The adiabatic
wind stagnation (central) temperature is shown by the dotted line. }
\label{fig1}
\end{figure}

Fig.~\ref{fig1} displays $T_{st}$ for different values of the adiabatic wind
terminal speed (or different ratios of energy to mass deposition rates) and
different metallicities of the thermalized plasma. The temperature at the
stagnation radius grows approximately as a quadratic function of $V_{A\infty}$
and remains always below the adiabatic value (dotted line  in Fig.~\ref{fig1}).
For practical purposes one can suggest that it does not depend on the plasma
abundance (see Fig.~\ref{fig1}). Note that for these calculations and
throughout the paper we have used the \citet*{1976ApJ...204..290R} equilibrium
cooling function tabulated and updated by \citet{1995MNRAS.275..143P}.

\subsection{The threshold mechanical luminosity}

In the case of a homogeneous stellar mass distribution, the equation
of mass conservation 
\begin{equation}
      \label{masscons}
\frac{1}{r^2} \der{}{r}\left(\rho_w u_w r^2\right) = q_m ,  
\end{equation}
can be easily integrated. At the star cluster edge this results in:
\begin{equation}
      \label{rhosc}
\rho_{SC} = \frac{q_m R_{SC}}{3 c_{SC}} \left(1 - 
            \frac{R^3_{st}}{R^3_{SC}}\right) .
\end{equation}

\begin{figure}
\centering
\resizebox{\columnwidth}{!}{\includegraphics{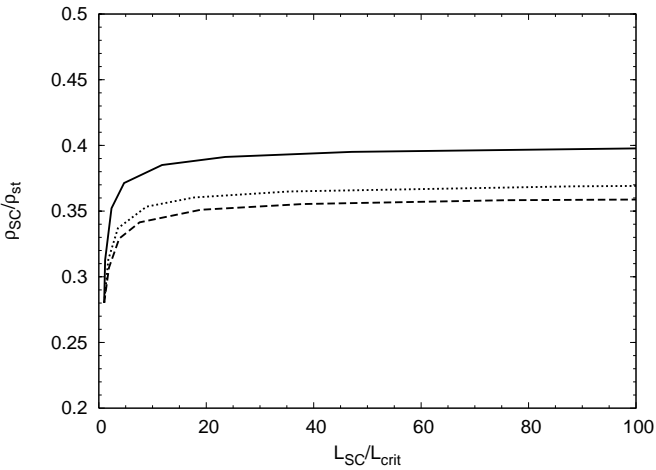}}
\caption{The ratio of the density at the star cluster surface to that at the 
stagnation radius in the catastrophic cooling regime. Solid, dotted and
dashed lines present semi-analytic results (see \citealt{2004ApJ...610..226S})
for $V_{A\infty}=1000$km s$^{-1}$, $V_{A\infty}=1500$km s$^{-1}$ and
$V_{A\infty}=2000$km s$^{-1}$, respectively.}
\label{fig2}
\end{figure}
As mentioned above, in the adiabatic case  the density at the 
star cluster surface scales linearly with that at the central stagnation 
radius. This is not exactly the case in the radiative solution. Nevertheless,
$\rho_{SC}/\rho_{st}$ remains  restricted into a very narrow interval,
$0.28 \le \rho_{SC}/\rho_{st} \le 0.42$, (see Fig.~\ref{fig2}), which
allows one to propose a simple analytic expression for the stagnation
radius in reasonable agreement with the results found  in semi-analytic and 
full numerical calculations.
Indeed, assuming by analogy with the adiabatic solution, that 
$\rho_{SC} = \alpha \rho_{st}$, where $\alpha$ is
a fiducial coefficient, one can combine Eqs.~(\ref{rhosc}) and
(\ref{rhoc}) to derive:
\begin{equation}
      \label{rstrsc}
\frac{R^3_{st}}{R^3_{SC}} = 1 - \left(\frac{L_{crit}}{L_{SC}}\right)^{1/2} ,
\end{equation}
where $L_{crit}$ is
\begin{equation}
      \label{lcrit}
L_{crit} = \frac{6 (\gamma-1) \pi \eta \alpha^2 \mu_{i}^2 R_{SC} V^4_{A\infty}}
           {(\gamma+1) \Lambda_{st}}
            \left(\frac{\eta V^2_{A\infty}}{2}  - \frac{c^2_{st}}{\gamma-1}\right) .
\end{equation}
In Eqs.~(\ref{rstrsc}) and (\ref{lcrit}) $\Lambda_{st}$ and $c_{st}$ are the  
cooling function value  and speed of sound at the stagnation radius, respectively.  To derive these equations 
we have used the adiabatic relation for the sound speed at the star 
cluster surface: $c^2_{SC} = (\gamma-1) V^2_{A\infty}/(\gamma+1)$.
When the mechanical luminosity of the cluster exceeds $L_{crit}$, the
stagnation radius detaches from the star cluster center, which
implies that $L_{crit}$ is the threshold luminosity which separates
clusters evolving in the catastrophic cooling (bimodal) regime from 
those evolving either in the quasi-adiabatic or in the radiative regimes,
with their stagnation radius located at the star cluster center. 
Note that for the latter cases  Eq.~(\ref{rstrsc}) is not valid.

\begin{figure}[t]
\centering
\resizebox{\columnwidth}{!}{\includegraphics{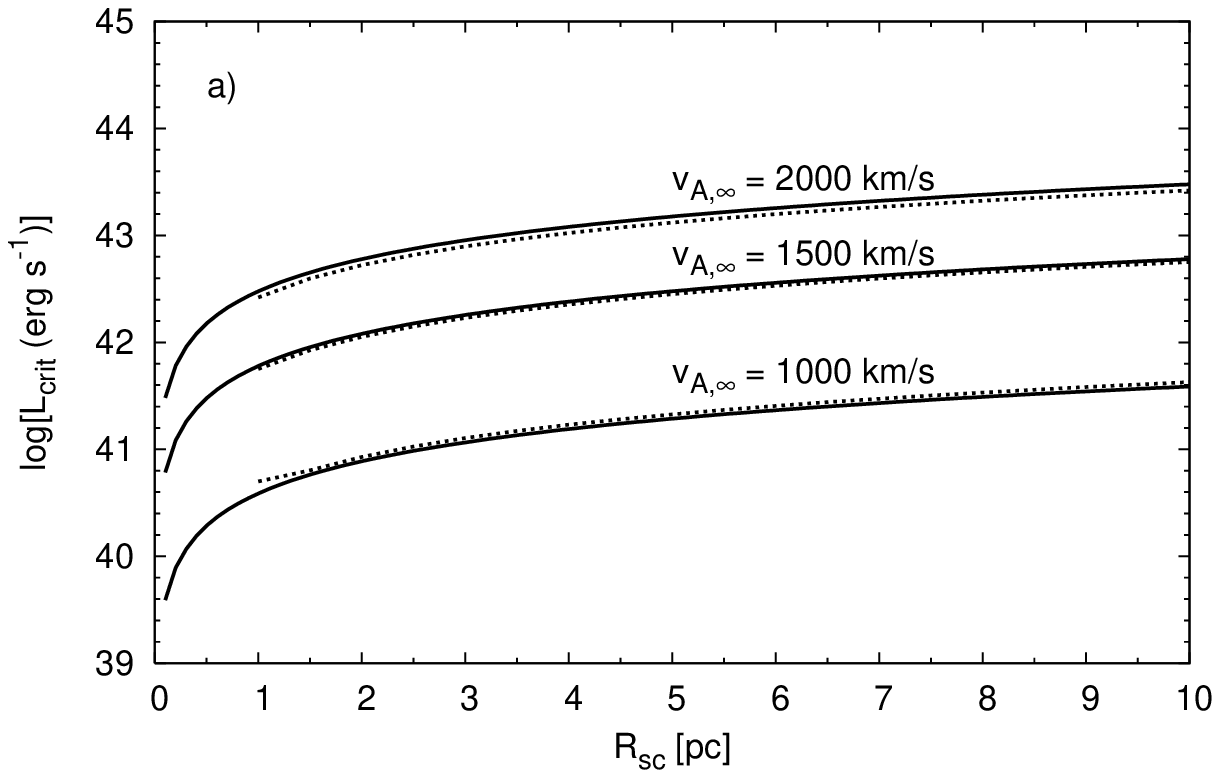}}
\resizebox{\columnwidth}{!}{\includegraphics{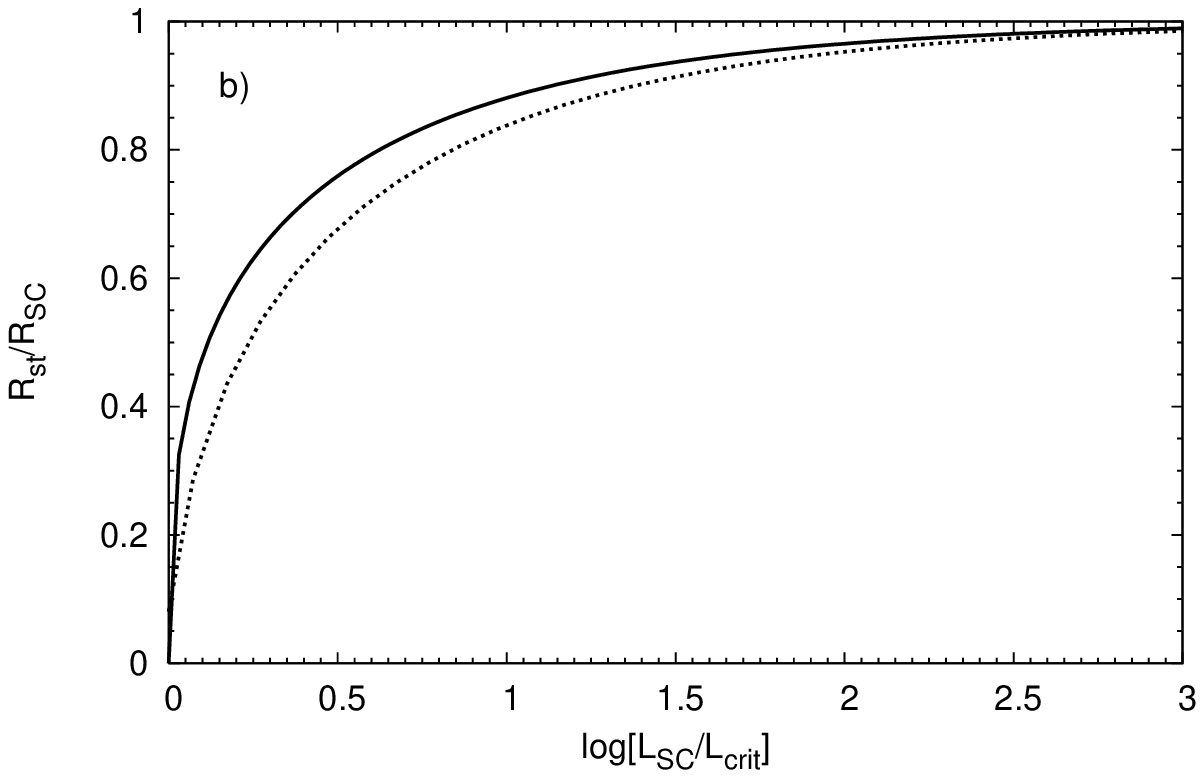}}
\resizebox{\columnwidth}{!}{\includegraphics{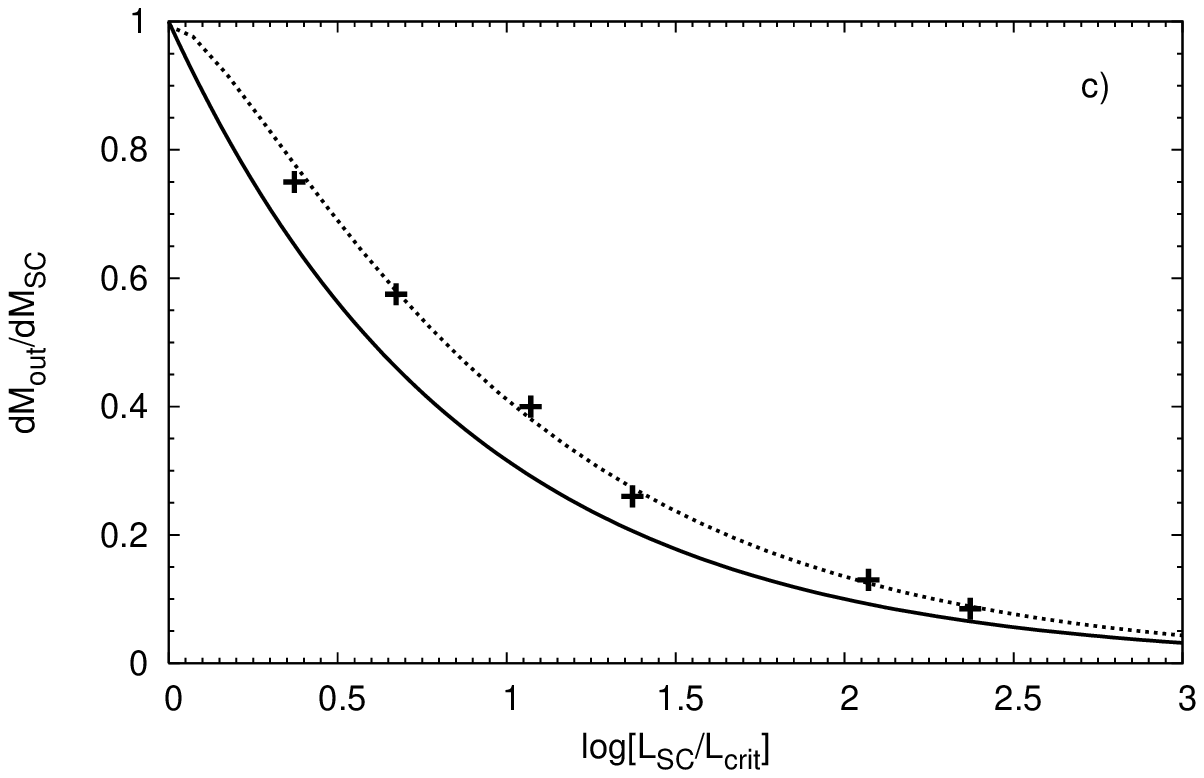}}
\caption{The comparison of analytic results with semi-analytic 
and full numerical calculations. Panel a) displays the critical luminosity 
calculated for three different values of the adiabatic 
wind terminal speed: $V_{A\infty} = 1000$ km s$^{-1}$, $V_{A\infty} = 1500$ 
km s$^{-1}$ and $V_{A\infty} = 2000$ km s$^{-1}$. 
Panel b) displays how the stagnation radius shifts from the center when
the mechanical luminosity exceeds the critical value. 
Panel c) shows the fraction of the matter supplied by supernovae and 
stellar winds that the cluster returns to the ISM of the host galaxy. In 
all panels solid lines present the analytic results.
The results from the semi-analytic and full numerical calculations are 
shown by the dotted line and by the cross symbols, respectively.
The fiducial coefficient $\alpha$ was set to $\alpha = 0.28$ in all 
calculations.}
\label{fig3}
\end{figure}
\subsection{The analytic formulation}

The analytic values for the threshold luminosity and the stagnation radius have
been compared with our semi-analytic results and full numerical calculations
(see Fig.~\ref{fig2} and \ref{fig3} in \citealt{2006astro.ph.12184T}). For that
we have used first Fig.~\ref{fig1} to obtain the temperature at the stagnation
radius for a given value of the adiabatic wind terminal speed, $V_{A\infty}$. We
then  use this temperature to calculate the sound speed, $c_{st}$, and the
corresponding cooling function value, $\Lambda_{st}$, and then use these values
to calculate the threshold luminosity and the stagnation radius from
Eqs.~(\ref{lcrit}) and (\ref{rstrsc}). The agreement between the approximate
analytic expressions and semi-analytic results is good if the fiducial
coefficient $\alpha = 0.28$ (see Fig.~\ref{fig3}).

Using Eq.~(\ref{rstrsc}) one can also estimate the fraction of 
the deposited matter that clusters evolving in the  
bimodal regime return to the ISM of their host galaxy:
\begin{equation}
      \label{Mout}
\frac{{\dot M}_{out}}{{\dot M}_{SC}} = \frac{R^3_{SC} - R^3_{st}}{R^3_{SC}} = 
     \left(\frac{L_{crit}}{L_{SC}}\right)^{1/2} \ ,
\end{equation}
This quantity  decreases monotonically with the star cluster mechanical 
luminosity (see Fig.~\ref{fig3}) as was also found in the 1D hydrodynamic
simulations performed by \citet{2006astro.ph.12184T}.

The analytic expression (\ref{Mout}) reproduces the numerical results
with an accuracy better than $\sim 25$\%. Minor  differences between the analytic
and the numerical outputs indicate that $\rho_{SC}/\rho_{st}$ deviates
from the assumed constant value.

\section{Heating efficiency}

Here we follow the \citet{1985Natur.317...44C} formulation supported by full
numerical calculations (see, for example \citealt{2000ApJ...536..896C};
\citealt{2004ApJ...604..662R}) and assume that the injected matter is
thermalized in situ  via random encounters of high velocity stellar winds and
supernovae ejecta. The efficiency of this process is an important, but poorly
known parameter \citep{2003MNRAS.339..280S}. This parameter has been discussed
by different groups (see, for example, \citealt{1998A&A...337..338B}; 
\citealt{2004A&A...424..817M} and
references therein) who argued for very different values of heating efficiency
that vary from 100\% to a few per cent. 

Note that different authors have different prescriptions for the physical
processes that may lead to an incomplete transformation of the deposited kinetic
energy. In our view, the high density of the sources may cause an important
fraction of the star cluster mechanical luminosity to be immediately radiated
away. This would happen locally  at the sites where the ejecta from nearby
stellar winds and supernovae would interact, inhibiting in this way the
possibility of having such energy    evenly spread throughout the cluster
volume. Turbulence may also absorb some fraction of the deposited energy. The
immediate consequence of such a sudden depletion of energy is to reduce the
thermal pressure gradient responsible for the acceleration of the injected
matter and thus the magnitude of the resultant  high velocity outflow. We have
incorporated all uncertainties related to this process into a parameter $\eta$
that represents the fraction of the deposited mechanical energy per unit time
which is finally evenly spread throughout the cluster volume and participates in
the thermal acceleration of the injected matter. The heating efficiency then
effectively reduces the value of the deposited energy, $L_{SC}$, but does not
affect the mass deposition rate, ${\dot M}_{SC}$.

\begin{figure}
\centering
\resizebox{\columnwidth}{!}{\includegraphics{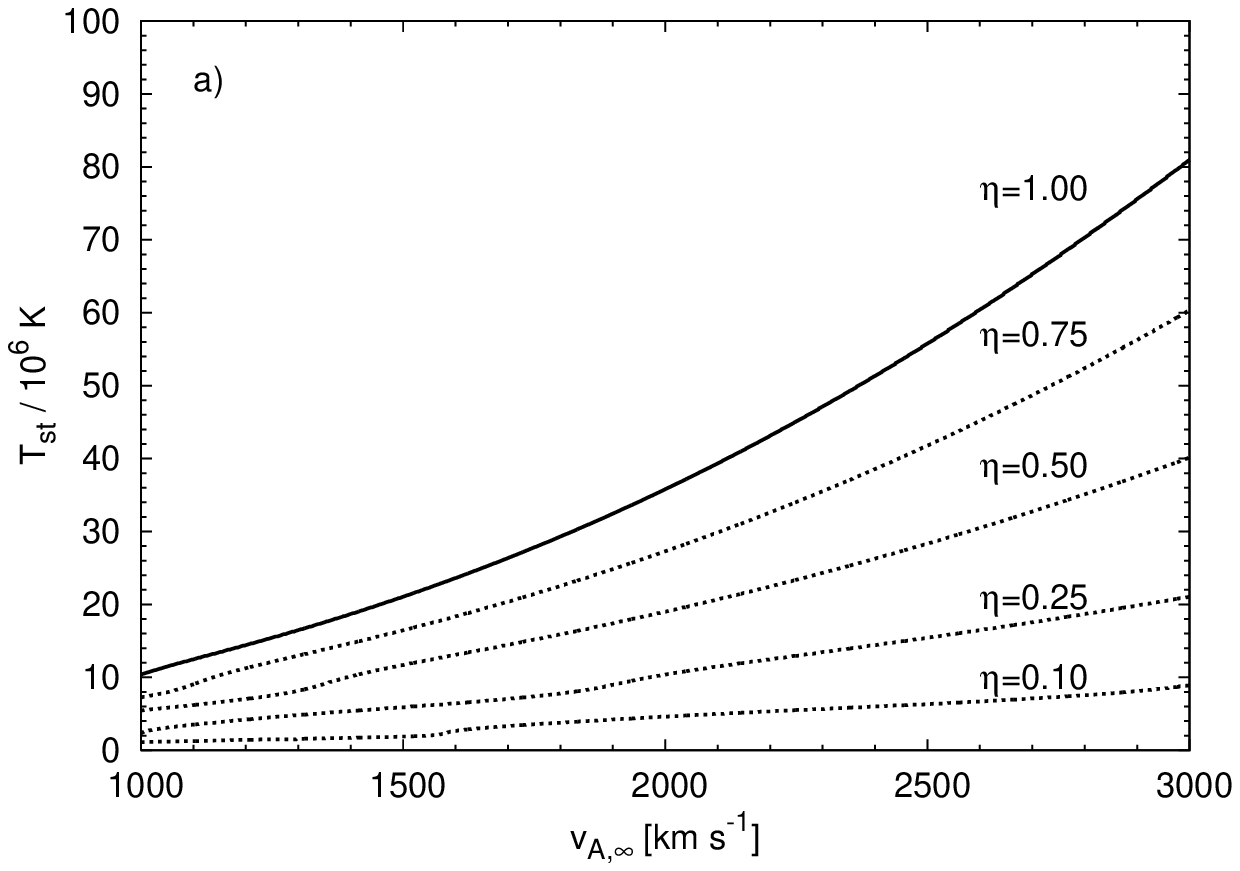}}
\resizebox{\columnwidth}{!}{\includegraphics{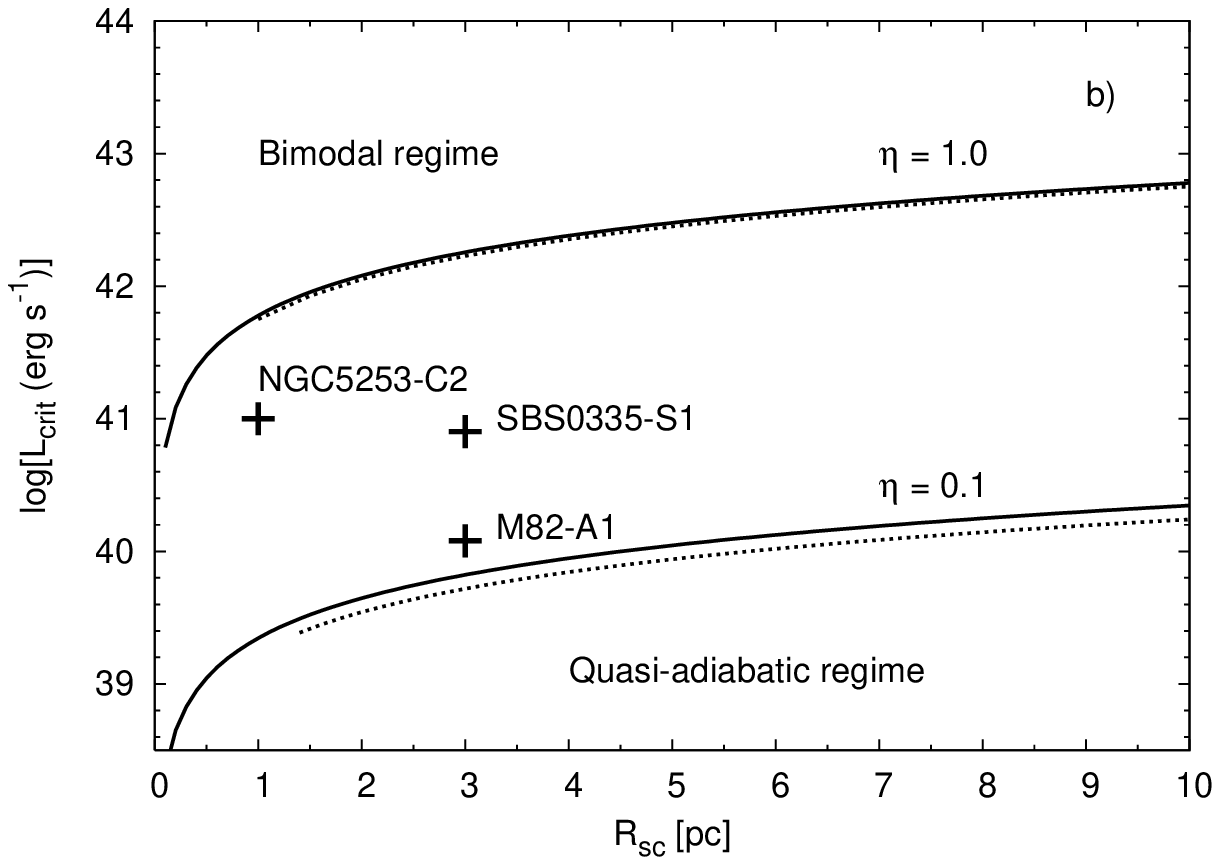}}
\caption{The impact of the heating efficiency. Panel a) displays the
temperature at the stagnation radius for different heating efficiencies,
$\eta$.  Panel b) displays the threshold mechanical luminosity for two 
heating efficiencies, $\eta = 1$ and $\eta = 0.1$. The analytic 
results (solid lines) are compared with the
semi-analytic calculations (see \cite{2006IAUS..237E..42S}) indicated by 
dotted lines.
The location of several massive SSCs
with respect to the threshold lines are indicated in panel b)
by cross symbols (see \citealt{2003Natur.423..621T,2004ApJ...602L..85T}; 
\citealt{2006ApJ...638..176T}; \citealt{2006MNRAS.370..513S}).
}
\label{fig4}
\end{figure}

Fig.~\ref{fig4} shows the impact of a reduced heating efficiency on the  plasma
temperature and how it drastically decreases the threshold luminosity. What
the most reasonable value of the heating efficiency in super star clusters is,
remains an open question in need of both detailed full hydrodynamical
calculations and careful analysis of the available observational data.
Nevertheless, the threshold mechanical luminosities calculated for a reasonable
range of heating efficiencies, $0.1 < \eta < 1$  (shown in Fig.~\ref{fig4}b),
suggest that many SSCs may be found well above the threshold line, as in
the case of the M82-A1 supercluster. See \citet{2006MNRAS.370..513S}, who, based
on the size of its associated HII region, estimated a heating efficiency $\eta <
0.1$. A critical revision of the available super star cluster data will be the
subject of a forthcoming communication.

\section{Conclusions}

We have provided a simple  analytic formulation that allows for an  easy
determination in the energy deposition rate vs cluster size diagram, of
the location of the  threshold mechanical luminosity that separates clusters
evolving in either of  the two possible hydrodynamic  solutions: the
normally assumed negative feedback solution in which all mass supplied by SNe
and individual stellar winds leaves the cluster as a cluster wind ($R_{st} =
0$) and the  bimodal solution ($0 < R_{st} < R_{SC}$) 
in which the reinserted matter  accumulates within the volume defined by $R_{st}$. 
Simple analytic  expressions,
accounting also for a reduced heating efficiency, allow one to calculate the
position of the stagnation radius and the fraction of the deposited matter that
clusters evolving in  the bimodal regime return to the ISM of their host
galaxy. Our results are in good agreement with our former semi-analytic and
numerical results.

\begin{acknowledgements}
We thank our anonymous referee for multiple comments and suggestions that
greatly improved the paper.  This study was partly supported by the
Institutional Research Plan AV10030501 of the Astronomical Institute, Academy of
Sciences of the Czech Republic and project LC06014 Center for Theoretical
Astrophysics. We also acknowledge support from Conacyt (M\'exico) grant 47534-F
and grants AYA2004-08260-C03-01 and AYA 2004-02703 from the Spanish Ministerio
de Educaci\'on y Ciencia.
\end{acknowledgements}

\bibliographystyle{aa}
\bibliography{winds2}

\end{document}